\documentclass[runningheads]{llncs}
\usepackage{graphicx}
\usepackage{amssymb}
\usepackage{xcolor}
\usepackage{epstopdf}
\usepackage{latexsym}
\usepackage{hyperref}
\usepackage{todonotes}
\usepackage{booktabs,dcolumn}
\usepackage{multirow}
\usepackage{amsmath,amssymb,amsfonts,bm}
\usepackage{mathtools} 
\usepackage{pifont}

\begin{document}
\title{Unsupervised Adaptive Implicit Neural Representation Learning for Scan-Specific MRI Reconstruction}
\titlerunning{Unsupervised Adaptive INR Learning for Scan-specific MRI Reconstruction}
%
\author{Junwei Yang\inst{1} \and
Pietro Li\`o\inst{1}}
\authorrunning{J. Yang et al.}
%

\institute{Department of Computer Science and Technology, University of Cambridge, Cambridge, United Kingdom \\
\email{\{jy406,pl219\}@cam.ac.uk}}
\maketitle              
\begin{abstract}
In recent studies on MRI reconstruction, advances have shown significant promise for further accelerating the MRI acquisition. Most state-of-the-art methods require a large amount of fully-sampled data to optimise reconstruction models, which is impractical and expensive under certain clinical settings. On the other hand, for unsupervised scan-specific reconstruction methods, overfitting is likely to happen due to insufficient supervision, while restrictions on acceleration rates and under-sampling patterns further limit their applicability. To this end, we propose an unsupervised, adaptive coarse-to-fine framework that enhances reconstruction quality without being constrained by the sparsity levels or patterns in under-sampling. The framework employs an implicit neural representation for scan-specific MRI reconstruction, learning a mapping from multi-dimensional coordinates to their corresponding signal intensities. Moreover, we integrate a novel learning strategy that progressively refines the use of acquired $k$-space signals for self-supervision. This approach effectively adjusts the proportion of supervising signals from unevenly distributed information across different frequency bands, thus mitigating the issue of overfitting while improving the overall reconstruction. Comprehensive evaluation on a public dataset, including both 2D and 3D data, has shown that our method outperforms current state-of-the-art scan-specific MRI reconstruction techniques, for up to 8-fold under-sampling.

\keywords{MRI reconstruction \and Implicit neural representation.}
\end{abstract}

\section{Introduction}
The technique of Magnetic Resonance Imaging (MRI) has been widely applied in clinical diagnostics, yet the prolonged scanning times, susceptibility to motion artefacts, and patient discomfort poses demands on accelerating the acquisition while preserving the imaging quality. Accelerated MRI methods, which learns to reconstruct from scans sampled at sub-Nyquist rates, have emerged as solutions to mitigate these issues by exploiting data redundancies, as seen in parallel imaging and compressed sensing MRI (CS-MRI) \cite{griswold2002generalized,lustig2007sparse,pruessmann1999sense}. However, these techniques can fail to generate high-fidelity images from highly sparsely-sampled ones \cite{sung2013high,yang2018admm}. Recently, deep learning approaches such as convolutional neural networks (CNNs) \cite{hyun2018deep,lee2017deep,qin2018convolutional,schlemper2017deep,seitzer2018adversarial,wang2016accelerating,yang2017dagan}, transformers \cite{huang2022swin,korkmaz2022unsupervised,zhou2023dsformer}, and diffusion models \cite{chung2022score,gungor2023adaptive,song2021solving} have shown great promise for accelerated MRI, particularly when sufficient supervised data is available, demonstrating superior reconstruction quality over conventional methods.

Given the high cost and impracticality of acquiring large volumes of fully-sampled data, unsupervised learning models that leverage only under-sampled data have been developed to benefit from reduced scan times \cite{cole2021fast,korkmaz2022unsupervised,yaman2020self,zhou2023dsformer}. Unlike common strategies that focus on exploiting the consistency in sampled regions, some models enhance reconstruction by partitioning measurements into subsets to maintain data consistency, but they generally require substantial data for optimising reconstruction models with complex architectures and may not be feasible in all clinical settings \cite{yaman2020self,zhou2023dsformer}. Scan-specific algorithms, on the other hand, offer a potential solution when it is impractical to obtain a set of under-sampled MR data \cite{akccakaya2019scan,griswold2002generalized,lustig2007sparse}. But their performance may drop significantly when the acceleration rate is extreme or the under-sampling pattern deployed is not well-designed for the methods to learn to interpolate.

In recent years, the application of Neural Radiance Field (NeRF) techniques to general \cite{mildenhall2022nerf,mildenhall2021nerf,weng2022humannerf} and MRI \cite{huang2023neural,shen2022nerp} reconstruction shows promise, though challenges in MRI reconstruction mainly include learning directly from $k$-space signals with straightforward architecture designs of implicit neural representation (INR) learning. The difficulty mainly arises from the sparse supervision, and the highly abstract regional correlation in the continuous representation in the frequency domain. 

Given the effectiveness of INR-based methods in reconstruction, an effective strategy of learning in $k$-space could further enhance INR-based reconstruction. For state-of-the-art methods, they typically either improve performance through consistency-based regularisation between the two domains \cite{eo2018kiki,zhou2020dudornet}, or transform the magnitude of $k$-space data to smoothen the highly-skewed distribution \cite{han2019k}, enabling the network to better exploit the raw $k$-space data. However, both approaches have only been evaluated in supervised learning settings, and might not be as effective without fully-sampled data for supervision. 

To overcome the limitations of existing scan-specific methods, we propose to improve the INR learning-based unsupervised reconstruction methods by adaptively learning from the highly imbalanced $k$-space data. Specifically, our strategy employs a fully-connected network to predict signal intensities from given coordinates in the image domain or $k$-space. Furthermore, we also propose a novel adaptive coarse-to-fine learning strategy that effectively reconstructs data by initially fitting low-frequency general information, then refining high-frequency details, regardless of under-sampling patterns or acceleration rates, allowing for broader applications of the proposed reconstruction framework.

We evaluated our proposed framework using a public multi-contrast MRI dataset, reconstructing 2D and the 3D under-sampled data, to demonstrate the capacity of our proposed method across different data representations commonly used in clinical settings. For under-sampling, we employed a non-uniform variable density pattern up to 8-fold acceleration. Our findings reveal the notable efficacy of the proposed techniques, which also outperform the state-of-the-art scan-specific reconstruction methods.

In summary, our key contributions are as threefold: 1) We propose an end-to-end scan-specific MRI reconstruction technique that improves the reconstruction; 2) We introduce a coarse-to-fine optimisation approach tailored for unsupervised learning, where the imbalanced information in $k$-space can be efficiently captured to mitigate overfitting; and 3) We conduct extensive experiments to evaluate the effectiveness of our proposed framework and compare with other state-of-the-art algorithms on under-sampled MRI data of various data representations.

\section{Related Work}
\subsection{INR Learning for MRI Reconstruction}
The success of neural representation learning has enabled new approaches for MRI reconstruction. Shen et al.~have proposed the pioneering application of INR learning in the field of medical imaging, particularly for reconstructing sparsely sampled images, including computed tomography (CT) and MRI scans. Their approach involves pre-training the model on a previous scan, which serves as the prior knowledge to improve the reconstruction of subsequent scans for the same patient \cite{shen2022nerp}. While the method is effective for scenarios requiring multiple visits, its utility is limited in cases where only a single session is needed.

Another effort in this field is the direct reconstruction of MRI data in $k$-space. A study conducted by Huang et al.~focused on dynamic cardiac cine MRI data, under-sampled using radial sampling patterns \cite{huang2023neural}. Despite showing promise due to the dense low-frequency sampling and unique coordinate representation, this applicability of this technique across varying under-sampling patterns remains questionable. Our preliminary studies using Cartesian under-sampling patterns revealed a significant drop in performance, which can restrict its clinical application in broader settings.

\subsection{Unsupervised MRI Reconstruction}
Unsupervised reconstruction techniques have been developed to exploit the partially sampled $k$-space data. An example is to use generative adversarial networks (GANs) to reconstruct images from under-sampled data \cite{cole2020unsupervised}. Additionally, several approaches have focused on exploiting the inherent consistency within under-sampled $k$-space data. For instance, the Self-Supervised learning via Data Under-sampling (SSDU) approach partitions original $k$-space data into two sets: one for data-consistency regularisation and the other for optimisation \cite{yaman2020self}.

A particularly innovative method, DSFormer, proposes a transformer-based architecture for multi-contrast MRI reconstruction \cite{zhou2023dsformer}. This approach divides under-sampled data into two non-overlapping subsets, with each subset occupying half of the sampled positions in $k$-space. Unsampled regions are filled with corresponding reference-contrast data values. These values are fed into the reconstruction network, and the consistency between the two resultant outputs provides additional supervision, alongside the conventional data-consistency loss.

\section{Method}
\label{sec:method}
\begin{figure}[t!]
    \centering
    \includegraphics[width=0.9\textwidth]{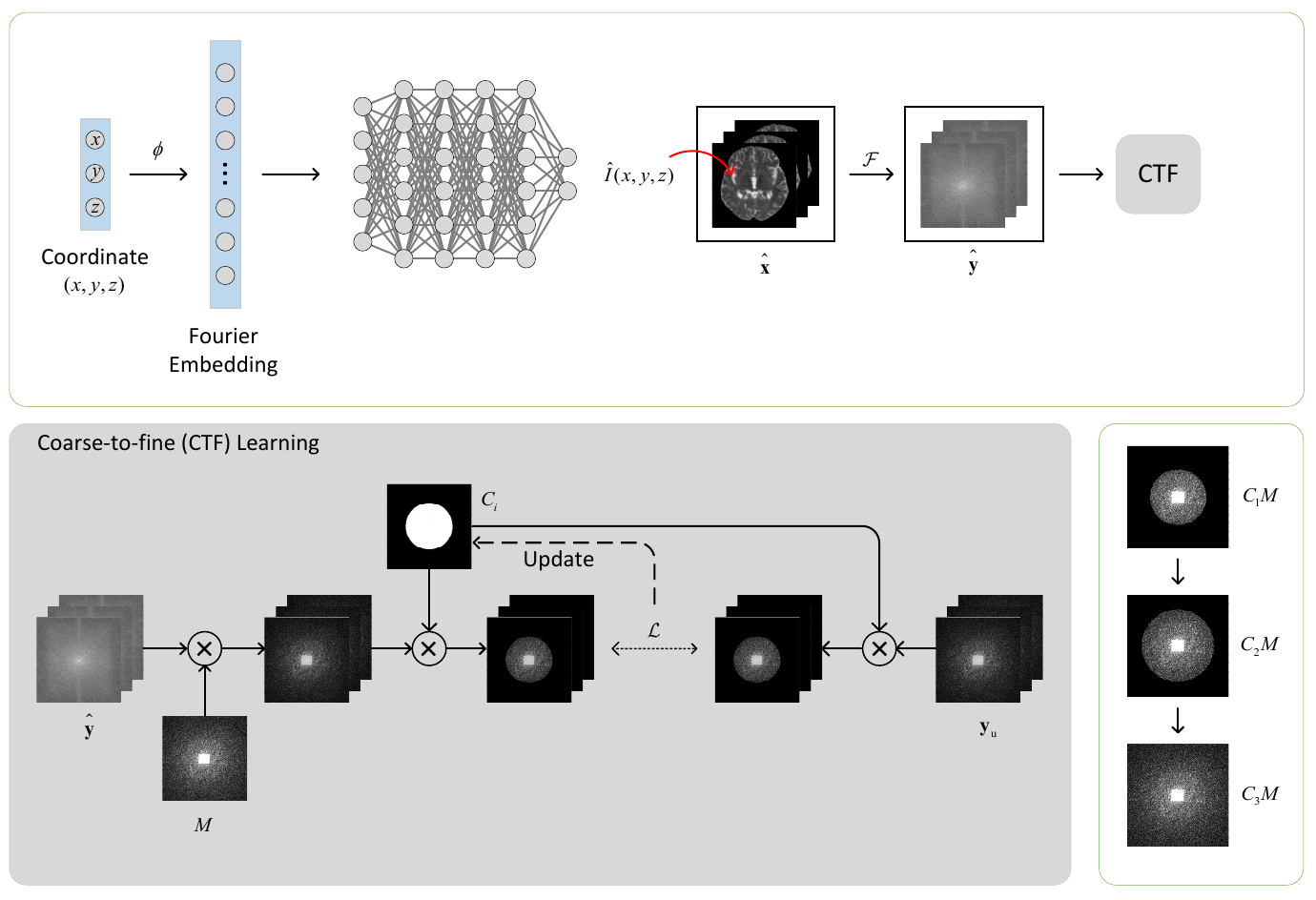}
    \caption{Illustration of our proposed framework. Top: The overall architecture of the proposed network, designed to learn implicit neural representation for reconstructing under-sampled MR images. Given spatial coordinates in the image domain or $k$-space, the network can generate the intensity values at the corresponding positions. Once the images are reconstructed by filling values in all spatial positions, their corresponding $k$-space data can be compared with the acquired data to optimise the network through the proposed coarse-to-fine learning strategy. Bottom left: The proposed coarse-to-fine learning strategy. The optimisation begins by focusing on the central low-frequency region of the $k$-space data, selected by a circular binary mask $C_1$. The initial optimisation is performed until convergence. Then, the data selection mask is expanded to include more positions from the outer $k$-space, and the optimisation is continued on the expanded $k$-space region. The iterative expansion and optimisation is repeated until all the acquired data is used for learning. Bottom right: An example of how the data selection mask evolves throughout the optimisation, where the $k$-space data is divided into three distinct frequency regions.}
    \label{fig:framework}
\end{figure}

Our methodology, as depicted in Fig.~\ref{fig:framework}, includes two main components. The first component is the Multi-Layer Perceptron (MLP) model for MRI reconstruction, which serves as the foundation of our approach. By leveraging the capabilities of a MLP, the implicit continuous representation can be effectively captured by the model directly from under-sampled MRI data. On the other hand, the second component is the coarse-to-fine adaptive supervision module, which is the novel aspect of our proposed framework, designed to dynamically adjust the granularity of supervision, in alignment with the progression of the optimisation process. By calibrating the level of detail in supervision, it allows the reconstruction network to more effectively exploit different levels of details based on sparsely under-sampled $k$-space data, from general to finer details, thus enhancing the overall quality of reconstructed images.

\subsection{Background and Problem Definition}
\label{subsec:overview_acc_mri}
To apply the implicit neural representation learning for reconstructing an under-sampled MR scan, we start with a coordinate in the image domain or $k$-space, denoted as $(x, y, z)$. A scan is represented as a set of intensity values at these positions. The complex-valued, under-sampled $k$-space data of the scan can be defined as $\textbf{y}_{\rm{u}} \in \mathbb{C}^{W \times H \times D}$, where $W$, $H$, $D$ are the width, height, and depth of the scan, respectively. To retrospectively under-sample the $k$-space data, a binary-valued under-sampling pattern, $M$, is applied to the fully-sampled data $\textbf{y} \in \mathbb{C}^{W \times H \times D}$, resulting in $\textbf{y}_{\rm{u}} = M \odot \textbf{y}$, with $\odot$ denoting element-wise multiplication. To obtain the corresponding image, $\textbf{x}$ from its original $k$-space data, the inverse Fourier transform, $\mathcal{F}^{-1}$, is applied as: $\textbf{x} = \mathcal{F}^{-1}(\textbf{y})$. For the problem of MRI reconstruction, given the under-sampled data, $\textbf{y}_{\rm{u}}$, a network $f_{\theta}$ parameterised by $\theta$ is designed to obtain the reconstructed data $\hat{\textbf{y}}$ by optimising the parameters in $f_{\theta}$.

\subsection{Implicit Neural Representation}
In INR, the network $f_{\theta}$ learns a mapping from a coordinate $d = (x, y, z)$ to the corresponding intensity in an MRI scan as:
\begin{equation}
I(d) = f_{\theta}(E(d)),
\end{equation}
where $I(d)$ is the intensity at the scan position $(x, y, z)$, and $E$ is an encoding function. Through optimisation, the network can thus learn a neural representation of the scan from sparsely sampled coordinate-intensity pairs, allowing it to infer and reconstruct unsampled positions.

To allow the subtle differences in magnitude of coordinates and high-frequency features to be better captured by the network, the Fourier encoding $E$ is applied to the coordinate first as:
\begin{equation}
E(d) = [\textnormal{sin}(2\pi A\phi(d)), \textnormal{cos}(2\pi A\phi(d))],
\end{equation}
where $A$ is a matrix randomly sampled from a zero-mean Gaussian distribution with the standard deviation of $\sigma$, and $\phi$ represents the coordinate normalisation operator. The matrix $A$ has dimensions of $N_e \times N_d$, with $N_e$ being the dimension of the encoded coordinate, and $N_d$ being the dimensionality of the coordinates $d$. The normalisation operator $\phi$ transforms the coordinate range for each dimension to lie between 0 and 1. Consequently, a coordinate of the size $N_d$ is mapped to a higher dimension of $2N_e$ after Fourier encoding, which can then be passed to the network $f_\theta$ as input for optimisation.

\subsection{MRI Reconstruction with INR}
\label{sec:mmri}
To learn the INR of a given scan, a MLP model can be used to reconstruct images, as depicted in the top of Fig.~\ref{fig:framework}. The intensity at each position $d$, denoted as $\hat{I}(d)$, is obtained from the optimised network as $\hat{I}(d) = f_\theta(E(d))$. By filling with the predicted intensity values across all positions, we can reconstruct the image $\hat{\textbf{x}}$ of the scan.

To optimise $f_\theta$, the objective function is defined for the predicted intensities. In our study, the L2 loss is employed for supervision, and the loss term is defined as:
\begin{equation}
    \mathcal{L} = ||M \odot \mathcal{F}(\hat{\textbf{x}}) - \textbf{y}_{\rm{u}}||_2^2.
\end{equation}

After optimisation, $f_\theta$ can infer intensities for unsampled positions. However, since the corresponding $k$-space value at sampled positions might not be perfectly predicted, the final reconstructed image $\hat{\textbf{x}}^{*}$ can be obtained by incorporating the data-consistency operation proposed in \cite{schlemper2017deep} to replace with the actual sampled values, ensuring the data consistency as:

\begin{equation}
    \hat{\textbf{x}}^{*} = \mathcal{F}^{-1}((1 - M) \odot \mathcal{F}(\hat{\textbf{x}}) + M \odot \textbf{y}_{\rm{u}}).
\end{equation}

\subsection{Coarse-to-fine Optimisation}
Since the supervision is provided in the form of $k$-space measurements at sampled positions, their highly imbalanced magnitude distribution makes optimisation challenging. This is particularly true for high-frequency regions with much smaller values that correspond to intricate details in the reconstructed images. Consequently, if all sampled values are used at the same time for supervision, the optimisation might predominantly focus on low-frequency regions due to their generally larger magnitude values. This can be detrimental for reconstructing finer details, which are essential for downstream applications such as diagnosis. To address this, we introduce a coarse-to-fine strategy, specifically designed for learning in $k$-space to accommodate the optimisation issue caused by imbalanced distribution. As the optimisation progresses and learns finer details, the addition of extra information can be helpful in mitigating the overfitting at each time of convergence.

The overall procedure is illustrated at the bottom of Fig.~\ref{fig:framework}. Initially, the network is supervised based on low-frequency $k$-space data. As the network is well trained with the low-frequency data, it is gradually exposed to data with higher frequency that is inherently more challenging to learn. The introduction of a hyper-parameter, step $S$, is important in this process as it determines the number of subsets into which the $k$-space data is divided. Specifically, a centred, circle-shaped binary mask $C_i$ with a defined radius to segment the $k$-space with a specific proportion of sampled points for the $i$-th step. Initially, the network is trained based on coarse, low-frequency information, with the initial mask $C_1$ capturing $(1 / S) \times 100\%$ of sampled points with the lowest frequency. Following convergence as described in Section \ref{sec:mmri}, the network is then exposed to information of relatively higher frequency. In practice, the radius of the mask $C_2$ is expanded to include a broader region, and $(2 / S) \times 100\%$ of the total sampled points are included. This iterative process continues until all the points are considered for supervision. 

\subsection{Model Details}
Within the INR learning-based framework, we employ a MLP to map coordinates to their corresponding intensity values. This MLP includes ten layers, where each layer, except the final one, has 256 neurons, uses the sine function as the activation function. This design principle has proven effective in learning the neural representation for both the image and its derivative, and we adhere to the proposed initialisations and values of hyper-parameters as proposed in \cite{sitzmann2020implicit}. For input, the size is matched to the dimensionality of the scanned image. The final layer outputs two-channel intensity values, representing the real and imaginary parts for the specified coordinates.

\section{Experiments}
\label{sec:experiments}

\subsection{Dataset}
Our evaluation was conducted on the dataset from the 2019 MICCAI Multi-modal Brain Tumor Segmentation (BraTS) Challenge \cite{menze2014multimodal}, which is a public brain MRI dataset. This dataset consists of 335 preprocessed brain MRI volumes for each contrast, acquired from multiple institutions. For each subject, we selected 15 central axial slices from both T2-weighted (T2w) and fluid-attenuated inversion recovery (FLAIR) sequences to create the volumetric data. Each slice was cropped to a dimension of $192 \times 192$ pixels to minimise background areas. The images were then normalised to have intensity values ranged between 0 and 1. We obtained the retrospective \textit{k}-space data by applying a Fourier transform to the images. In our experiments, the T2w and FLAIR contrasts were chosen to be reconstructed. The dataset includes volumes from 15 subjects, with 4 used for validation and the remaining for evaluation. These volumes were used to evaluate our proposed method on the 3D data representation, while individual axial slices from the same volumes were used for evaluating 2D data representation performance.

\subsection{Implementation Details}
In the MLP network architecture used for MRI reconstruction, values of several hyper-parameters need to be defined. For the Fourier encoding, the Fourier feature dimension was set to 256, with a standard deviation of 1 for the variable $A$. In each iteration, the $k$-space values selected by our coarse-to-fine strategy were considered for supervision, as the Fourier transform is needed for calculating the loss on predicted images. The network is trained for up to 10,000 iterations using the Adam optimiser \cite{kingma2014adam} until convergence was observed. The optimiser parameters $\beta_1$ and $\beta_2$ were set to 0.5 and 0.999, respectively. The learning rates were 0.0001 for 2D data and 0.00001 for 3D data. All training was conducted on an NVIDIA A100 graphics card.

The optimal hyper-parameter values were determined through a grid search on the validation set, with the number of coarse-to-fine steps, $S$, set to 3. After optimisation, the network can predict the intensities at unsampled coordinates, to allow for image reconstruction.

\subsection{Evaluation}
To quantitatively evaluate our proposed method, we employed peak signal-to-noise ratio (PSNR) and structural similarity index measure (SSIM). PSNR calculations were performed on the magnitude images, comparing the reconstructed data $\hat{\textbf{x}}$ with the ground-truth data $\textbf{x}$. On the other hand, SSIM evaluations followed the same principles of comparison, and the implementation is based on the standard protocol defined in \cite{wang2004image}.

For evaluation, reconstructed images were compared against fully-sampled ground-truth images. We used a Poisson-disc pattern for retrospective under-sampling at acceleration rates of $4\times$ and $8\times$, while a $32 \times 32$ calibration region is applied. To achieve the desired acceleration rates, binary search was used to fine-tune the density slope as described in \cite{bridson2007fast}.

In terms of baseline methods, we included state-of-the-art scan-specific reconstruction algorithms, including the compressed sensing-based approaches. Additionally, $l1$-norm regularisations in total variation and wavelet domains are applied to provide extra constraints. The weights of the regularisation terms were fine-tuned on the validation set based on the grid search, with a selected value of 0.1 for both regularisations.

Scan-specific $k$-space reconstruction methods, such as GRAPPA and RAKI \cite{akccakaya2019scan,griswold2002generalized}, are popular in clinical settings. However, their limitations in handling certain under-sampling patterns caused by the dependency on regional information from sampled positions make them not applicable for certain scenarios. Consequently, including these methods as baselines would not be practical in our study, since they will fail in situations with unsampled regions according to the under-sampling patterns we deployed, which is a challenge that our proposed method addresses.

\section{Results}

\begin{table*}[t!]
\caption{Results of MRI reconstruction with different data representations (2D and 3D). The contrasts to be reconstructed include T2w and FLAIR, and acceleration rates include $4\times$ and $8\times$. CTF represents the proposed coarse-to-fine learning strategies, respectively.}
\centering
\resizebox{\columnwidth}{!}{
\begin{tabular}{lcllllllll}
\hline
\multicolumn{2}{l}{}                        & \multicolumn{4}{c}{T2w}                                                                              & \multicolumn{4}{c}{FLAIR}                                                                         \\ \cline{3-10} 
\multicolumn{2}{l}{Method (2D)} & \multicolumn{2}{c}{4$\times$}                       & \multicolumn{2}{c}{8$\times$}                       & \multicolumn{2}{c}{4$\times$}                       & \multicolumn{2}{c}{8$\times$}                       \\ \cline{3-10} 
\multicolumn{2}{l}{}                        & \multicolumn{1}{c}{PSNR} & \multicolumn{1}{c}{SSIM} & \multicolumn{1}{c}{PSNR} & \multicolumn{1}{c}{SSIM} & \multicolumn{1}{c}{PSNR} & \multicolumn{1}{c}{SSIM} & \multicolumn{1}{c}{PSNR} & \multicolumn{1}{c}{SSIM} \\ \hline
             & CTF         &                          &                          &                          &                          &                          &                          &                          &                          \\ \cline{1-2}
Ours           & \ding{51}           & \bf{42.82$\pm$2.40}                    & \bf{0.982$\pm$0.0017}                    & \bf{35.57$\pm$1.00}                    & \bf{0.956$\pm$0.013}                    & \bf{43.92$\pm$3.72}                    & \bf{0.986$\pm$0.015}                    & \bf{39.01$\pm$3.09}                    & \bf{0.974$\pm$0.027}                    \\
Ours            & \ding{55}           & 41.61$\pm$2.54                    & 0.978$\pm$0.017                    & 34.78$\pm$1.25                    & 0.945$\pm$0.020                    & 42.92$\pm$3.39                    & 0.982$\pm$0.020                    & 37.89$\pm$3.98                    & 0.958$\pm$0.039                    \\ \hline
\multicolumn{2}{l}{CS (TV)}              & 33.44$\pm$1.89                    & 0.966$\pm$0.0090                   & 29.35$\pm$1.92                    & 0.939$\pm$0.018                    & 33.00$\pm$2.95                    & 0.969$\pm$0.0074                    & 28.19$\pm$3.16                    & 0.943$\pm$0.017                    \\
\multicolumn{2}{l}{CS (Wavelet)}         & 35.93$\pm$1.85                    & 0.969$\pm$0.0052                    & 30.53$\pm$1.72                    & 0.937$\pm$0.014                    & 35.16$\pm$3.11                    & 0.969$\pm$0.0074                    & 30.08$\pm$2.97                    & 0.937$\pm$0.019                    \\ \hline
\end{tabular}}

\centering
\resizebox{\columnwidth}{!}{
\begin{tabular}{llllllllll}
\hline
\multicolumn{2}{l}{}                        & \multicolumn{4}{c}{T2w}                                                                              & \multicolumn{4}{c}{FLAIR}                                                                         \\ \cline{3-10} 
\multicolumn{2}{l}{Method (3D)} & \multicolumn{2}{c}{4$\times$}                       & \multicolumn{2}{c}{8$\times$}                       & \multicolumn{2}{c}{4$\times$}                       & \multicolumn{2}{c}{8$\times$}                       \\ \cline{3-10} 
\multicolumn{2}{l}{}                        & \multicolumn{1}{c}{PSNR} & \multicolumn{1}{c}{SSIM} & \multicolumn{1}{c}{PSNR} & \multicolumn{1}{c}{SSIM} & \multicolumn{1}{c}{PSNR} & \multicolumn{1}{c}{SSIM} & \multicolumn{1}{c}{PSNR} & \multicolumn{1}{c}{SSIM} \\ \hline
                   & CTF         &                          &                          &                          &                          &                          &                          &                          &                          \\ \cline{1-2}
Ours                    & \ding{51}           & \bf{40.92$\pm$1.03}                    & \bf{0.975$\pm$0.00083}                    & \bf{34.69$\pm$1.70}                    & \bf{0.955$\pm$0.0035}                    & \bf{41.55$\pm$2.62}                    & \bf{0.975$\pm$0.0024}                    & \bf{35.75$\pm$1.37}                    & \bf{0.957$\pm$0.0066}                    \\
Ours                     & \ding{55}           & 38.30$\pm$1.10                    & 0.971$\pm$0.0028                    & 32.12$\pm$0.91                    & 0.940$\pm$0.0078                    & 38.95$\pm$1.74                    & 0.968$\pm$0.0051                    & 35.43$\pm$2.10                    & 0.956$\pm$0.0060                    \\ \hline
\multicolumn{2}{l}{CS (TV)}              & 30.22$\pm$2.01                    & 0.960$\pm$0.012                    & 28.22$\pm$2.13                    & 0.933$\pm$0.022                    & 28.85$\pm$2.74                    & 0.945$\pm$0.028                    & 25.46$\pm$3.99                    & 0.920$\pm$0.036                    \\
\multicolumn{2}{l}{CS (Wavelet)}         & 32.19$\pm$1.74                   & 0.970$\pm$0.0071                    & 29.30$\pm$1.74                    & 0.942$\pm$0.015                    & 29.57$\pm$4.37                    & 0.947$\pm$0.031                    & 26.47$\pm$4.27                    & 0.920$\pm$0.032                    \\ \hline
\end{tabular}}

\label{tab:main_res}
\end{table*}

\subsection{2D MRI Reconstruction}
We first evaluated our proposed method on 2D MRI data, with the quantitative performance of the reconstruction model presented in the first table of Table \ref{tab:main_res}, in which the metrics were calculated for each slice individually. The results demonstrate that our method significantly outperforms existing scan-specific reconstruction algorithms at both $4\times$ and $8\times$ acceleration rates for T2w and FLAIR contrasts. Notably, our proposed novel coarse-to-fine optimisation strategy contributed to the reconstruction, which yielded noticeable improvements over the conventional optimisation strategy.

\begin{figure}[t]
    \centering
    \includegraphics[width=0.97\textwidth]{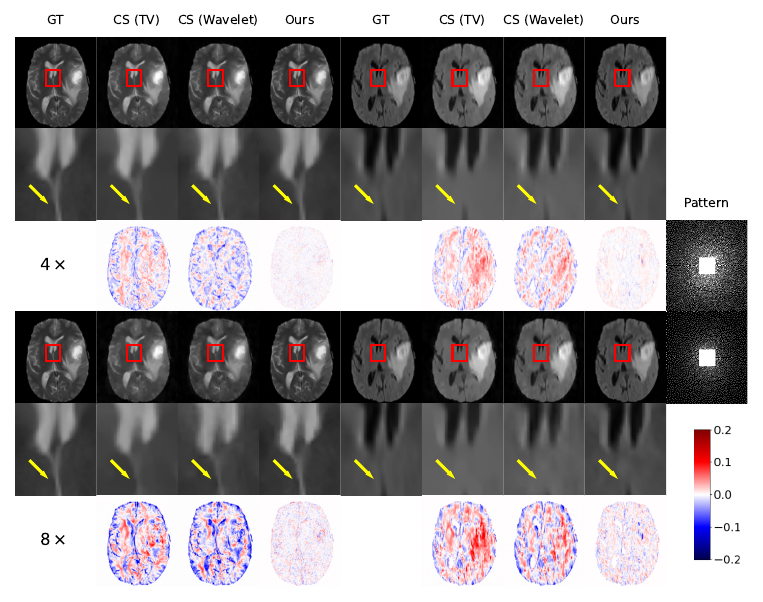}
    \caption{Reconstruction results on 2D MRI data under the acceleration rates of $4\times$ and $8\times$. For each acceleration rate, the first row displays the slice, while the second row zooms in on the highlighted red region, and the third row presents the respective error maps. The corresponding under-sampling patterns are illustrated in the last column. }
    \label{fig:res_2d}
\end{figure}

Furthermore, selected reconstruction results are visualised in Fig.~\ref{fig:res_2d}. The reconstructed images and corresponding error maps demonstrate the capability of our proposed method in capturing finer anatomical details, and it is consistent across both contrasts and all evaluated acceleration rates, indicating the superiority of our algorithm in reconstructing high-quality MR images from sparsely sampled $k$-space data.

\subsection{3D MRI Reconstruction}
We further expanded our evaluation to reconstruct 3D whole-brain volumes on the same dataset. The network architecture used was the same as the one for 2D image reconstruction, with the only difference being the dimensionality of the input coordinates. For 3D reconstruction, a 3D Fourier transform was applied to derive the corresponding $k$-space data from the predicted intensity values.

The quantitative results are reported in the second table of Table \ref{tab:main_res} for both contrasts. These results are consistent with the observations from the 2D data evaluation, further confirming the efficacy of our proposed method. Similarly, significant performance improvements can be observed when comparing our approach to existing state-of-the-art reconstruction techniques, while the proposed coarse-to-fine learning strategy is effective in enhancing the performance.

\begin{figure}[t]
    \centering
    \includegraphics[width=0.97\textwidth]{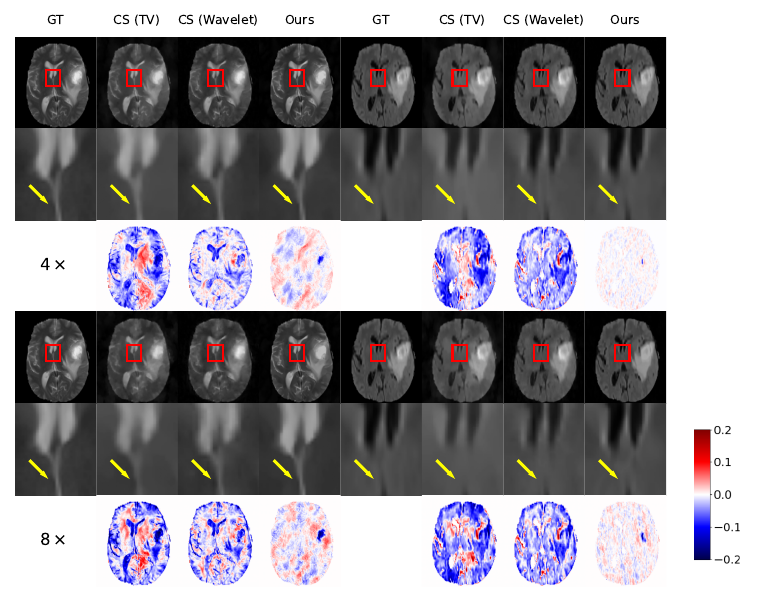}
    \caption{Reconstruction results on 3D MRI data under the acceleration rates of $4\times$ and $8\times$. For each acceleration rate, the first row displays the slice, while the second row zooms in on the highlighted red region, and the third row presents the respective error maps.}
    \label{fig:res_3d}
\end{figure}

The reconstruction images for 3D data are displayed in Fig.~\ref{fig:res_3d}. For comparison purposes, the same slices selected for the 2D data evaluation are extracted. Even with sparsely under-sampled data, our proposed method demonstrates precision in reconstructing whole-brain volumes, successfully capturing detailed structures that other methods struggle to reproduce. This visual representation, combined with the quantitative data, indicates the superior performance of our proposed reconstruction method in a 3D context.

\subsection{Comparison between Different Inputs}
The evaluation of our method, as summarised in the two tables of Table \ref{tab:main_res}, includes both 2D and 3D data representations as inputs for the MLP model. To ensure consistency and fairness in our comparison, we used the same ground-truth data for both representations. Specifically, for the 3D data, we extracted axial slices from the volumetric data for computing the metrics.

The results indicate that our proposed method achieves the best performance with 2D data, followed by 3D data. This higher performance in 2D reconstructions can be attributed to the relatively simpler representation of 2D data, which makes the optimisation easier for the model. However, this advantage comes with a drawback in efficiency, as it requires separate training for each slice. As for 3D data reconstruction, while the task is more challenging due to higher variations in anatomy across slices, it offers better efficiency. Specifically, a model trained on 3D data can reconstruct the entire volume, without the need for training on each slice.

The reconstructed slices displayed in Fig.\ref{fig:res_2d} and Fig.\ref{fig:res_3d} visually confirms these findings. The 2D representation typically yields reconstructions of higher quality, capturing finer details more effectively. On the other hand, direct reconstructions from 3D volumes tend to appear less sharp or accurate. This observation reflects the trade-offs between data complexity, model performance, and operational efficiency in MRI reconstruction of different data representations.

\begin{figure}[t]
    \centering
    \includegraphics[width=0.57\textwidth]{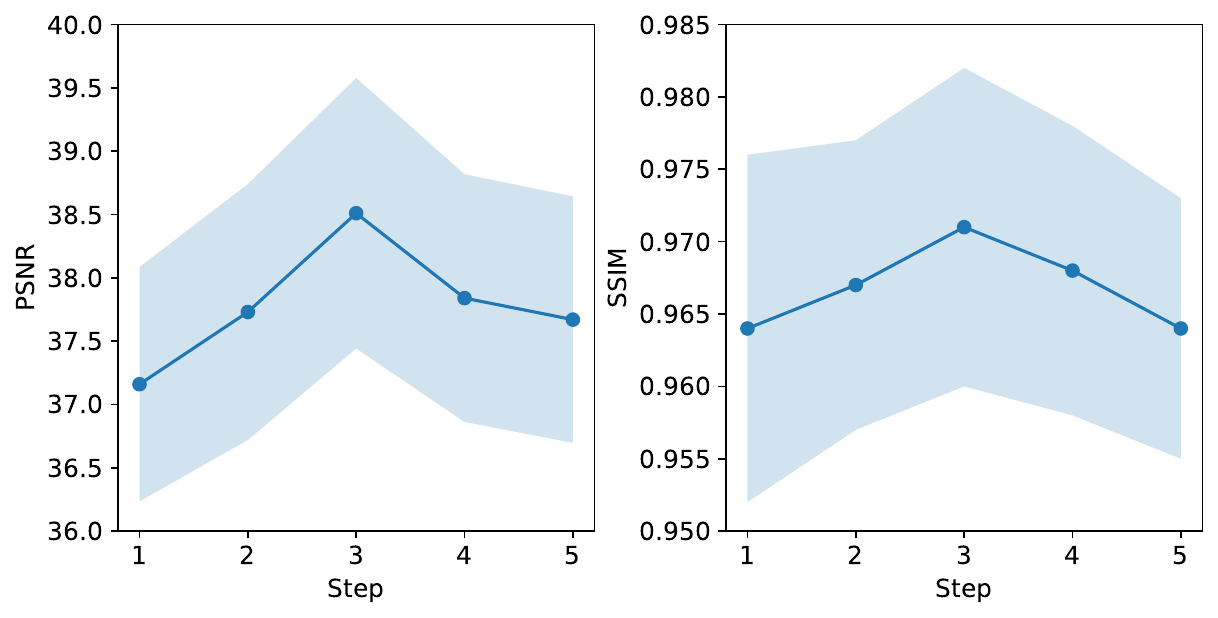} \\
    \includegraphics[width=0.45\textwidth]{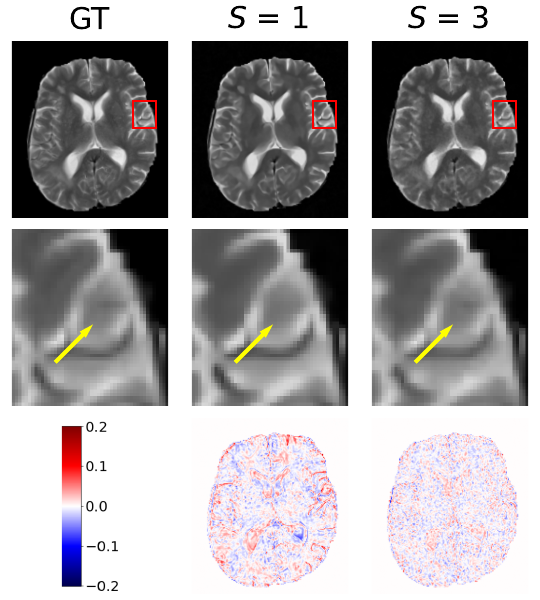}
    \caption{Quantitative (top) and visualised (bottom) reconstruction results with varied values of $S$, in which T2w serves as the contrast to be reconstructed. }
    \label{fig:influence_s}
\end{figure}

\subsection{Influence of Coarse-to-fine Learning}
\label{sec:eml}
To better understand the impact of the coarse-to-fine learning strategy by varying the associated hyper-parameter, extensive experiments were conducted on the dataset with different values of the step $S$. Specifically, the hyper-parameter $S$ determines the granularity of the optimisation process, denoting the number of steps in which additional supervision of increasing level of details is introduced. When $S = 1$, it implies supervision using all sampled $k$-space values at the same time. On the other hand, higher $S$ values result in a partitioned approach, dividing the sampled measurements into several distinct sets, with each step including data of increasing details.

In our experiments, the validation set was used to evaluate the proposed framework with varying $S$ values up to the division into five sets. As depicted in Fig.~\ref{fig:influence_s}, the reconstruction performance was evaluated across a various $S$ values. Initially when $S = 1$, where the coarse-to-fine strategy is not introduced, the model is considered as the baseline. As the data was divided into an increasing number of sets, a noticeable improvement in reconstruction quality was observed. This improvement is likely due to the capacity of the model of sequentially focusing on different frequency bands in $k$-space, allowing it to capture a broader range of details.

However, the performance starts to decline when $S$ exceeds 3, suggesting that while the strategy can be beneficial, the model might not be able to generalise effectively across various frequency bands with excessive divisions.

In the visualised examples, with reconstructions at $S = 3$ displaying optimal quality, effectively capturing both coarse and fine details, the observation suggests a balance between the granularity and the capacity of the model to effectively learn from the $k$-space data.

\begin{figure}[t]
    \centering
    \includegraphics[width=0.87\textwidth]{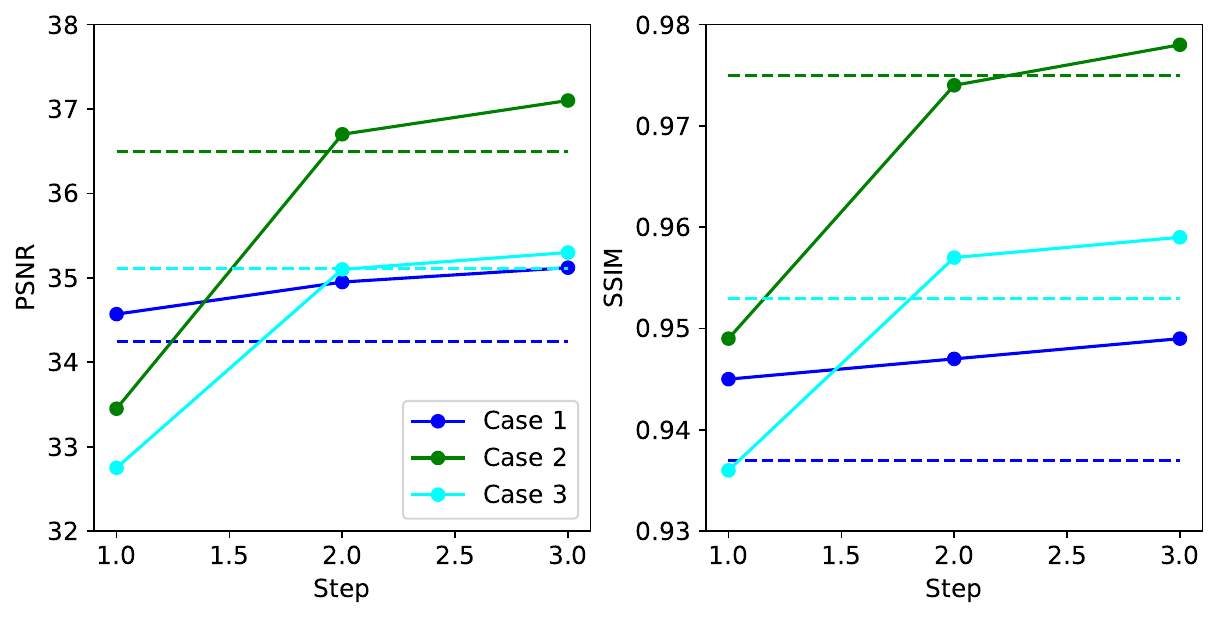}
    \caption{The reconstruction results during the optimisation of three examples, in which T2w serves as the contrast to be reconstructed. Results across all fineness levels in a single run are reported for each case, and the horizontal dashed line with the same colour represents the performance without the coarse-to-fine learning strategy. }
    \label{fig:single_run}
\end{figure}

To further demonstrate the impact of supervision with data of different frequencies, Fig.~\ref{fig:single_run} shows the reconstruction performance of three examples during the progress of training with the optimal combination of hyper-parameter values. The quality of the reconstructed images improves progressively with the addition of data with more details. Notably, the performance generally starts to outperform that of the method without the coarse-to-fine strategy after the first step, indicating its effectiveness.

\section{Discussion and Conclusion}
\label{sec:discussion}
In this paper, we address the problem of scan-specific MRI reconstruction, focusing on improving the reconstruction quality of under-sampled data by exploiting the statistical properties across different frequency bands in $k$-space. Our proposed framework is based on the application of INR learning, which can effectively capture the complex representation within sparse $k$-space measurements. To mitigate overfitting and deal with the complexities of learning in $k$-space, we introduce a novel coarse-to-fine optimisation strategy. This approach enables the model to initially capture broader structural features of the scan before focusing on learning finer details. Beyond improving MRI reconstruction, this strategy can also be potentially extended to broader applications where full supervision is available.

We conducted extensive experiments using a public multi-contrast MRI dataset and considered various data representations to evaluate the efficacy of our proposed method, especially using diverse data representations commonly encountered in clinical settings. The results reveal that our proposed method outperforms state-of-the-art algorithms, achieving up to an 8-fold acceleration with variable-density under-sampling patterns. The integration of our coarse-to-fine strategy was proven effective in improving reconstruction performance across all evaluated data representations.

During our preliminary research, we examined alternative approaches for our coarse-to-fine strategy. One involved optimising the model with each disjoint set separately, while the other used multiple MLPs with each focusing on specific frequency bands. Both approaches, however, yielded unsatisfactory results. The former led to a loss of structural information due to a shift in focus to higher frequencies corresponding to details, while the latter struggled to accurately capture high-frequency details due to the insignificant difference within high-frequency information.

Nevertheless, our INR-based approach has its limitations. Its primary drawback is the requirement for individual training for each subject, which restricts its generalisability. Current studies aim to enhance the applicability of INR methods through the following strategies: 1) Improving generalisation by incorporating additional input features \cite{wu2022arbitrary}; and 2) Increasing computational efficiency to accelerate the optimisation \cite{muller2022instant,reiser2021kilonerf}. Given the flexibility of our framework, it can be adapted to incorporate these improvements, leveraging the full potential of INR-based learning in MRI reconstruction tasks. Another challenge is to adapt the model to handle an additional coil dimension in the input for multi-coil image reconstruction. While a direct application of our current framework shows competitive performance, according to our preliminary results, there is still great potential for further improvements by effectively utilising the correlation across coil images.

\bibliographystyle{splncs04}
\bibliography{lncs}

\end{document}